\documentclass[twoside,twocolumn,9pt]{article}
\usepackage{extsizes}
\usepackage[super,sort&compress,comma]{natbib} 
\usepackage[version=3]{mhchem}
\usepackage[left=1.5cm, right=1.5cm, top=1.785cm, bottom=2.0cm]{geometry}
\usepackage{balance}
\usepackage{times,mathptmx}
\usepackage{sectsty}
\usepackage{graphicx} 
\usepackage{lastpage}
\usepackage[format=plain,justification=justified,singlelinecheck=false,font={stretch=1.125,small,sf},labelfont=bf,labelsep=space]{caption}
\usepackage{float}
\usepackage{fancyhdr}
\usepackage{fnpos}
\usepackage[english]{babel}
\addto{\captionsenglish}{%
  
}
\usepackage{array}
\usepackage{droidsans}
\usepackage{charter}
\usepackage[T1]{fontenc}
\usepackage[usenames,dvipsnames]{xcolor}
\usepackage{setspace}
\usepackage[compact]{titlesec}
\usepackage{hyperref}
\usepackage{dblfloatfix}

\definecolor{cream}{RGB}{222,217,201}

\begin{document}

\pagestyle{fancy}
\thispagestyle{plain}


\makeFNbottom
\makeatletter
\renewcommand\LARGE{\@setfontsize\LARGE{15pt}{17}}
\renewcommand\Large{\@setfontsize\Large{12pt}{14}}
\renewcommand\large{\@setfontsize\large{10pt}{12}}
\renewcommand\footnotesize{\@setfontsize\footnotesize{7pt}{10}}
\makeatother

\renewcommand{\thefootnote}{\fnsymbol{footnote}}
\renewcommand\footnoterule{\vspace*{1pt}%
\color{cream}\hrule width 3.5in height 0.4pt \color{black}\vspace*{5pt}} 
\setcounter{secnumdepth}{5}

\makeatletter 
\renewcommand\@biblabel[1]{#1}            
\renewcommand\@makefntext[1]%
{\noindent\makebox[0pt][r]{\@thefnmark\,}#1}
\makeatother 
\renewcommand{\figurename}{\small{Fig.}~}
\sectionfont{\sffamily\Large}
\subsectionfont{\normalsize}
\subsubsectionfont{\bf}
\setstretch{1.125} 
\setlength{\skip\footins}{0.8cm}
\setlength{\footnotesep}{0.25cm}
\setlength{\jot}{10pt}
\titlespacing*{\section}{0pt}{4pt}{4pt}
\titlespacing*{\subsection}{0pt}{15pt}{1pt}

\fancyfoot{}
\fancyfoot[RO]{\footnotesize{\sffamily{ ~\textbar  \hspace{2pt}\thepage}}}
\fancyfoot[LE]{\footnotesize{\sffamily{\thepage~\textbar\hspace{3.45cm} }}}
\fancyhead{}
\renewcommand{\headrulewidth}{0pt} 
\renewcommand{\footrulewidth}{0pt}
\setlength{\arrayrulewidth}{1pt}
\setlength{\columnsep}{6.5mm}
\setlength\bibsep{1pt}

\makeatletter 
\newlength{\figrulesep} 
\setlength{\figrulesep}{0.5\textfloatsep} 

\newcommand{\topfigrule}{\vspace*{-1pt}%
\noindent{\color{cream}\rule[-\figrulesep]{\columnwidth}{1.5pt}} }

\newcommand{\botfigrule}{\vspace*{-2pt}%
\noindent{\color{cream}\rule[\figrulesep]{\columnwidth}{1.5pt}} }

\newcommand{\dblfigrule}{\vspace*{-1pt}%
\noindent{\color{cream}\rule[-\figrulesep]{\textwidth}{1.5pt}} }

\makeatother

\twocolumn[
  \begin{@twocolumnfalse}
\vspace{3cm}
\sffamily
\begin{tabular}{m{4.5cm} p{13.5cm} }

& \noindent\LARGE{\textbf{Probing the stability and magnetic properties of magnetosome chains in freeze-dried magnetotactic bacteria}} \\
\vspace{0.3cm} & \vspace{0.3cm} \\

 & \noindent\large{Philipp Bender,$^{\ast}$\textit{$^{a}$} Lourdes Marcano,\textit{$^{b}$}\textit{$^{c}$} I\~naki Orue,\textit{$^{d}$} Diego Alba Venero,\textit{$^{e}$} Dirk Honecker,\textit{$^{f\ddag}$} Luis Fern\'andez Barqu\'in,\textit{$^{g}$} Alicia Muela,\textit{$^{h}$}\textit{$^{i}$} and M. Luisa Fdez-Gubieda,\textit{$^{j}$}\textit{$^{h}$}} \\

\\
& \noindent\normalsize{\textit{Magnetospirillum gryphiswaldense} biosynthesize high quality magnetite nanoparticles, called magnetosomes, and arrange them into a chain that behaves like a magnetic compass.
Here we perform magnetometry and polarized small-angle neutron scattering (SANS) experiments on a powder of freeze-dried and immobilized \textit{M. gryphiswaldense}. 
We confirm that the individual magnetosomes are single-domain nanoparticles and that an alignment of the particle moments along the magnetic field direction occurs exclusively by an internal, coherent rotation.
Our magnetometry results of the bacteria powder indicate an absence of dipolar interactions between the particle chains and a dominant uniaxial magnetic anisotropy.
Finally, we can verify by SANS that the chain structure within the immobilized, freeze-dried bacteria is preserved also after application of large magnetic fields up to 1\,T.} \\

\end{tabular}

 \end{@twocolumnfalse} \vspace{0.6cm}

  ]

\renewcommand*\rmdefault{bch}\normalfont\upshape
\rmfamily
\section*{}
\vspace{-1cm}


\footnotetext{\textit{$^{a}$~Physics and Materials Science Research Unit, University of Luxembourg, 1511 Luxembourg, Grand Duchy of Luxembourg. Fax: +352 46 66 44 36521; Tel: +352 46 66 44 6521; E-mail: philipp.bender@uni.lu}}
\footnotetext{\textit{$^{b}$~Helmholtz-Zentrum Berlin f\"ur Materialien und Energie, 12489 Berlin, Germany.}}
\footnotetext{\textit{$^{c}$~Dpto. Electricidad y Electrónica, Universidad del País Vasco – UPV/EHU,
48940 Leioa, Spain.}}
\footnotetext{\textit{$^{d}$~SGIker, Universidad del País Vasco – UPV/EHU, 48940 Leioa, Spain.}}
\footnotetext{\textit{$^{e}$~ISIS Neutron and Muon Facility, Rutherford Appleton Laboratory,
Chilton, OX11 0QX, United Kingdom.}}
\footnotetext{\textit{$^{f}$~Large Scale Structures group, Institut Laue-Langevin, 38042 Grenoble, France.}}
\footnotetext{\textit{$^{g}$~CITIMAC, Universidad de Cantabria, 39005 Santander, Spain.}}
\footnotetext{\textit{$^{h}$~BCMaterials, Basque Center for Materials, Applications and Nanostructures, UPV/EHU Science.}}
\footnotetext{\textit{$^{i}$~Dpto. Immunologia, Microbiologia y Parasitologia, Universidad del País Vasco – UPV/EHU, 48940 Leioa, Spain.}}
\footnotetext{\textit{$^{j}$~Dpto. Electricidad y Electrónica, Universidad del País Vasco – UPV/EHU,
48940 Leioa, Spain.}}

\footnotetext{\ddag~ Present address: Physics and Materials Science Research Unit, University of Luxembourg, 1511 Luxembourg, Grand Duchy of Luxembourg.}



\section{Introduction}
Magnetotactic bacteria are microorganisms that are able to align and navigate along geomagnetic fields thanks to the presence of one or more chains of magnetic nanoparticles with high chemical purity synthesized in their interior (i.e., magnetosomes) \cite{Bazylinski2004,khalil2013closed}.
\textit{Magnetospirillum gryphiswaldense} for example contains a variable number of cuboctahedral magnetite ($\mathrm{Fe_3O_4}$) magnetosomes with a mean diameter of 40\,nm arranged in a chain \cite{Fdez-Gubieda2013}. 
This arrangement results from the interplay between the magnetic dipolar interaction, between nearest magnetite crystals, and a complex lipid/protein-based architecture that conform the cytoskeleton \cite{komeili2006magnetosomes,toro2019mamy}.
A previous study of this strain has shown that the magnetosomes arrange in helical-like shaped chains due to the tilting of the individual magnetic moments of the magnetosomes with respect to the chain axis \cite{Orue2018}.
This natural magnetic arrangement can be considered a prototype of a 1D magnetic nanoarchitecture and has motivated various studies to synthesize similar particles \cite{lee2011magnetosome,guardia2012water,lak2018fe2+,castellanos2019outstanding} and 1D structures \cite{martinez2013learning,sturm2019magnetic,andreu2019anisotropic}.

In general, linear 1D assemblies of magnetic nanoparticles have received considerable interest in various fields including micromechanical sensors \cite{yuan2011one,wang2011magnetic}, microfluidics \cite{cebers2005flexible}, micro-swimmers \cite{dreyfus2005microscopic} and also fundamental science \cite{kiani2015elastic}.
Different techniques exist to synthesize 1D nanoparticle chains \cite{martinez2013learning,sturm2019magnetic,wolf2005pattern,korth2006polymer,nakata2008chains} however, thanks to the high biological control imposed in the synthesis of the magnetosomes, magnetotactic bacteria produce 1D nanostructures with high reproducibility and quality \cite{prozorov2013novel}.
Due to their exceptional properties magnetotactic bacteria, such as \textit{M. gryphiswaldense}, remain highly investigated, motivating several experimental studies where their structural \cite{hoell2004nanostructure,bender2019using,rosenfeldt2019probing} and magnetic properties \cite{Marcano2017,marcano2018magnetic,gandia2019unlocking} were evaluated.
In \citet{koernig2014probing} and \citet{blondeau2018magnetic} it was observed that already small magnetic fields (i.e. small torques) of around 30\,mT can be sufficient to break-up the particle chains within alive but immobilized bacteria.

In this work we investigate the response of freeze-dried magnetotactic bacteria to large magnetic fields up to 1\,T.
We use a combination of DC magnetometry and small-angle neutron scattering (SANS) to determine the magnetic properties of the individual magnetosomes and the chains.
From the polarized SANS results we can gain additionally insight into the nanostructure of the 1D chains, enabling us to test their mechanical stability in high fields, i.e. high magnetic torques.

\section{Experimental section}

\textit{M. gryphiswaldense} strain MSR-1 (DMSZ 6631) was grown in a flask standard medium (FSM).\cite{Heyen2003} 
The culture was kept in three-fourths full bottles at 28\,$^\circ$C without agitation. 
After 120\,h, when the magnetosomes were well-formed \cite{Fdez-Gubieda2013}, the bacteria cells were collected and fixed in 2\% glutaraldehyde. 
After repeated washings with distilled water, the bacteria were freeze-dried, resulting in a powder sample with random orientation of the particle chains, i.e., a structurally and magnetically isotropic sample.
Additionally, we grew and freeze-dried bacteria without magnetosomes \citep{Fdez-Gubieda2013}, which were measured by SANS as a background sample to subtract from the magnetotactic bacteria signal.

To characterize the bacteria, we first performed transmission electron microscopy (TEM, JEOL JEM-1400Plus) on unstained bacteria adsorbed onto carbon-coated copper grids, and used ImageJ \cite{Schneider2012} for the image processing and analysis.

For the magnetization measurements the bacteria powder was introduced in a gelatin capsule and investigated with a superconducting quantum interference device magnetometer (Quantum Design MPMS-3) in DC mode.
The bacteria powder was dense enough to prevent rotation or any other physical movement of the bacteria even at large magnetic fields.
We collected isothermal remanence magnetization (IRM) and direct current demagnetization (DCD) curves, as well as an isothermal magnetization loop at room-temperature. 
The IRM curve was obtained by measuring the remanence from the initially demagnetized state by taking the sample through successive minor loops from 0 to 1\,T. 
In a similar manner, the DCD curve was obtained by a progressive demagnetization of the initially saturated sample, and the isothermal magnetization loop was measured between 1 and -1\,T in continuous mode. 
The diamagnetic contributions from the bacteria and sample holder were corrected.
Additionally, we measured first-order reversal curves (FORC) with a customized vibrating sample magnetometer between a saturating field of 0.1 T and increasing reversal fields, until reaching finally the major loop.
The resulting FORC diagrams were generated from a total of 80 minor loops according to \citet{Roberts2000}.

The polarized SANS (SANSPOL) experiment of the bacteria powder was conducted at room temperature with the Larmor instrument at ISIS neutron and muon source, Rutherford Appleton Laboratory.
The powder of the randomly oriented bacteria was enclosed in an aluminium sample holder.
It can be noted, that we observed no changes between repeated field-dependent measurements under the same conditions, which strongly indicates that the bacteria did not physically move during the experiment.
The magnetic field $\textbf{H}$ was applied perpendicular to the incoming neutron beam ($ \textbf{k}||\textbf{e}_x\perp\textbf{H}||\textbf{e}_z$).
The resulting SANS cross-sections can be written in case of an isotropic magnetization distribution in $\widetilde{M}_y$ as \cite{honecker2010longitudinal,michels2011small}
\begin{align}
I^{\pm}(\textbf{q})\propto&|\widetilde{N}|^2 + b_\mathrm{h}^2\left(  |\widetilde{M}_x|^2 + |\widetilde{M}_y|^2\mathrm{cos}^2\Theta + |\widetilde{M}_z|^2\mathrm{sin}^2\Theta\right)\nonumber\\
& \mp b_\mathrm{h}(\widetilde{N}\widetilde{M}_z^*+\widetilde{N}^*\widetilde{M}_z)\mathrm{sin}^2\Theta\label{eq4},
\end{align}
where the index $+/-$ indicates the polarization of the incoming beam, i.e. spin-up or spin-down.
Here $\Theta$ is the angle between the scattering vector $\textbf{q}=(0,q_y,q_z)$ and the magnetic field $\textbf{H}$, and $b_\mathrm{h}=2.7\cdot10^{-15}\,\mathrm{m}/\mu_\mathrm{B}$, where $\mu_\mathrm{B}$ is the Bohr magneton. 
Moreover, $\widetilde{N}(\textbf{q})$ and $\widetilde{M}_{x,y,z}(\textbf{q})$ denote the Fourier transforms of the nuclear scattering length density and of the magnetization in the $x$-, $y$- and $z$-directions, respectively.
From the measured SANSPOL intensities, the nuclear-magnetic cross-terms $I_\mathrm{cross}(q)\propto(\widetilde{N}\widetilde{M}_z^*+\widetilde{N}^*\widetilde{M}_z)$ can be determined as a function of the applied field from the sector perpendicular to $\textbf{H}$ of the exclusively polarisation dependent cross section (i.e. the residual scattering pattern) $I^{-}(\textbf{q})-I^{+}(\textbf{q})=I_\mathrm{cross}(q)\mathrm{sin}^2\Theta$.
Analysis of the cross-term is in particular useful for systems where the nuclear scattering dominates the magnetic scattering, which is usually the case e.g. for iron oxide nanoparticles \cite{hoell2004nanostructure,disch2012quantitative,bender2018relating,muhlbauer2019magnetic}.
Furthermore, we measured the SANS signal of the empty bacteria as background measurement and subtracted it from the SANSPOL intensities $I^{\pm}(\textbf{q})$ to be able to determine the nuclear scattering of the magnetosome chains.

\section{Results and  Discussion}

\begin{figure}[t]
\centering  
\includegraphics[width=1\columnwidth]{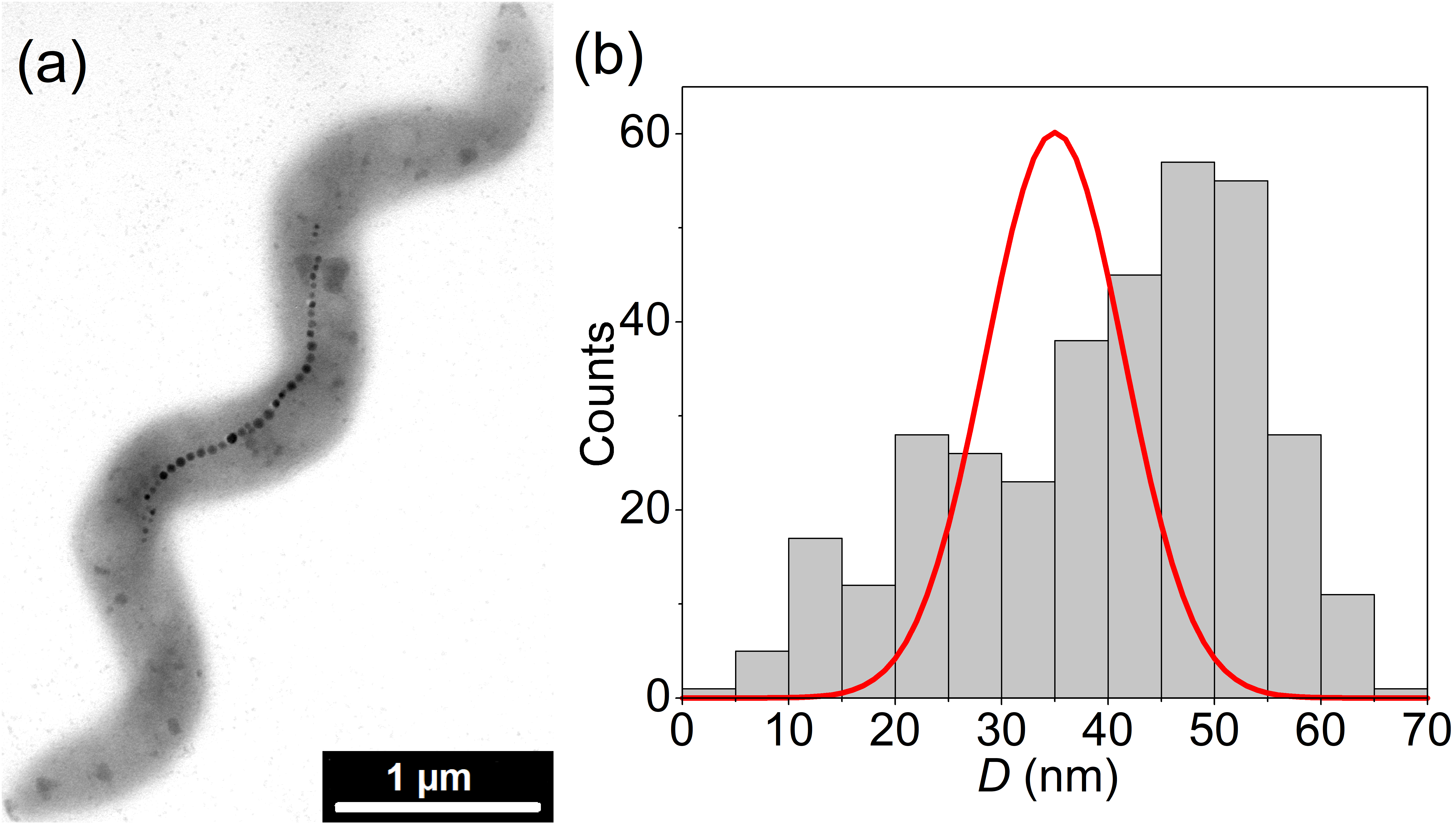}
\caption{\label{Fig1}(a) TEM image of a \textit{M. gryphiswaldense} with a chain of 35 particles, and (b) the size-histogram of the particles ($N=347$). The red line is the number-weighted size distribution (in arbitrary units) which we determined by fitting the SANS data under assumption of spherical particle shape.}
\end{figure}

Fig.\,\ref{Fig1}(a) shows a typical TEM image of a \textit{M. gryphiswaldense} containing a chain of magnetosomes. 
The observed chain is composed of uniformly-sized nanoparticles except for those at both ends of the chain, which are slightly smaller \cite{Orue2018}. 
This results in a broadening of the size histogram (Fig.\,\ref{Fig1}(b)), whose mean value is $\langle D \rangle = 40$\,nm.
Typically, the number of particles per cell is around 15 to 40, with a center-to-center distance between neighboring particles of about 50-60\,nm \cite{Orue2018}.

A shown in \citet{Fdez-Gubieda2013}, the magnetosomes synthesized by \textit{M. gryphiswaldense} are nearly perfect magnetite.
Using the bulk values of magnetite \cite{coey2010magnetism} for the exchange constant $A=7\,\mathrm{pJ/m}$, the first-order anisotropy constant $K_1=-13\,\mathrm{kJ/m^3}$, and the saturation magnetization $M_\mathrm{S}=0.48\,\mathrm{MA/m}$,  the critical single-domain size $D_\mathrm{SD}=72\sqrt{AK}/(\mu_0M_\mathrm{S}^2)$ (i.e., the size above which domain formation is energetically favorable \cite{skomski2003nanomagnetics}) can be calculated to $D_\mathrm{SD}\approx 75\,\mathrm{nm}$.
This is significantly above the mean size of the magnetosomes and thus we can assume that at first approximation they are single-domain particles.

\begin{figure}[t]
\centering  
\includegraphics[width=1\columnwidth]{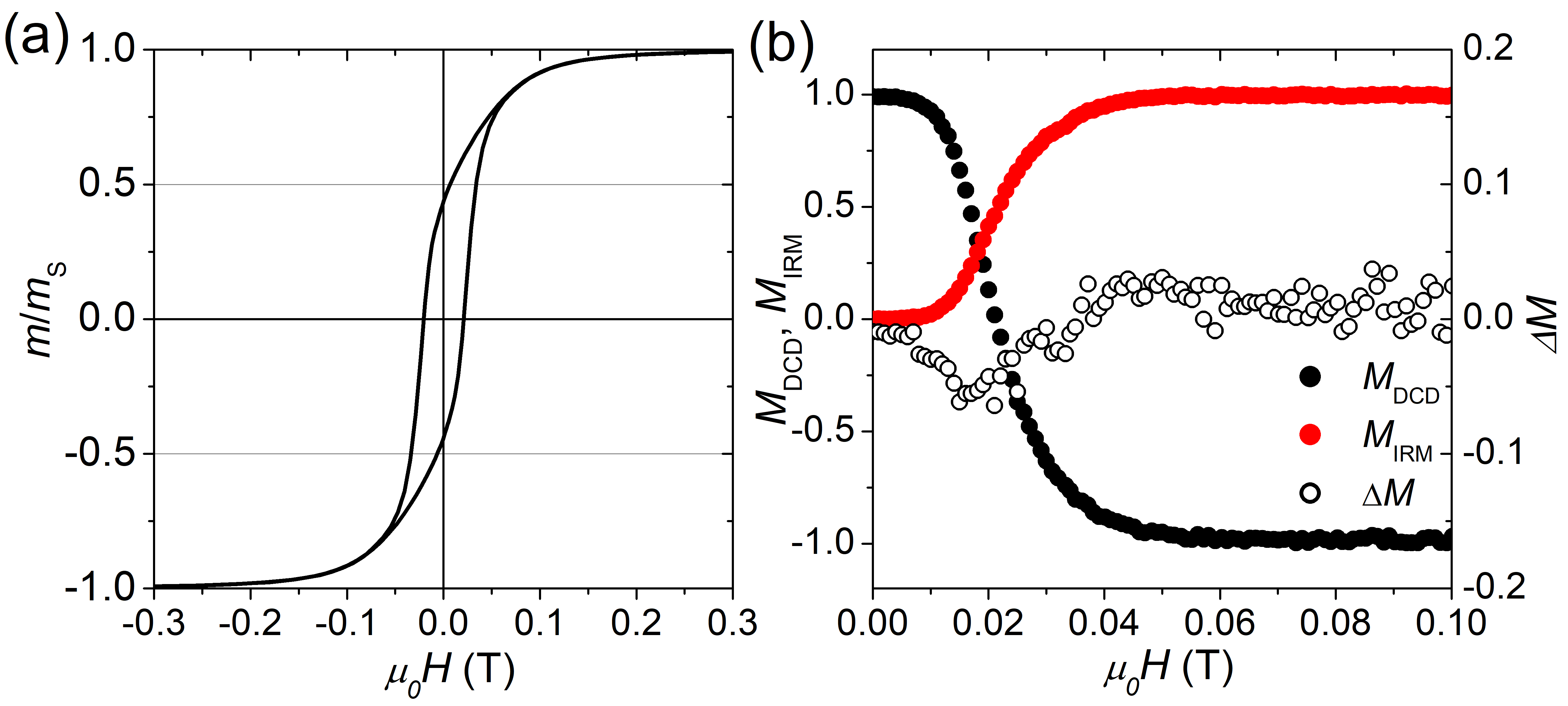}
\caption{\label{Fig2}Isothermal, room-temperature magnetization measurements of the freeze-dried bacteria powder: (a) Normalized magnetization curve. (b) Normalized remanence curves $M_\mathrm{DCD}$ and $M_\mathrm{IRM}$, and the modified Henkel plot $\Delta M$ (Eq.\,\ref{eq:2}).} 
\end{figure}

\begin{figure*}[ht]
\centering  
\includegraphics[width=2\columnwidth]{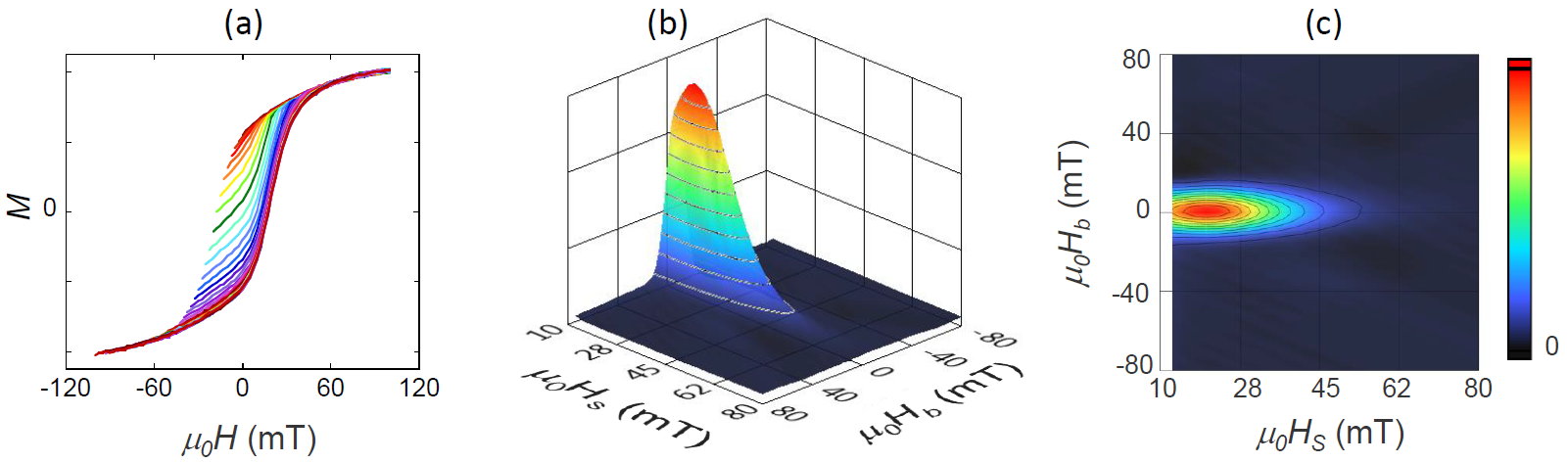}
\caption{\label{Fig3}(a) Set of 80 room-temperature FORC measurements in arbitrary units, and (b) the resulting 3D FORC diagram  and (c) the contour plot projected in the $(H_\mathrm{S},H_\mathrm{B})$ plane.} 
\end{figure*}

\begin{figure*}[hb]
\centering  
\includegraphics[width=2\columnwidth]{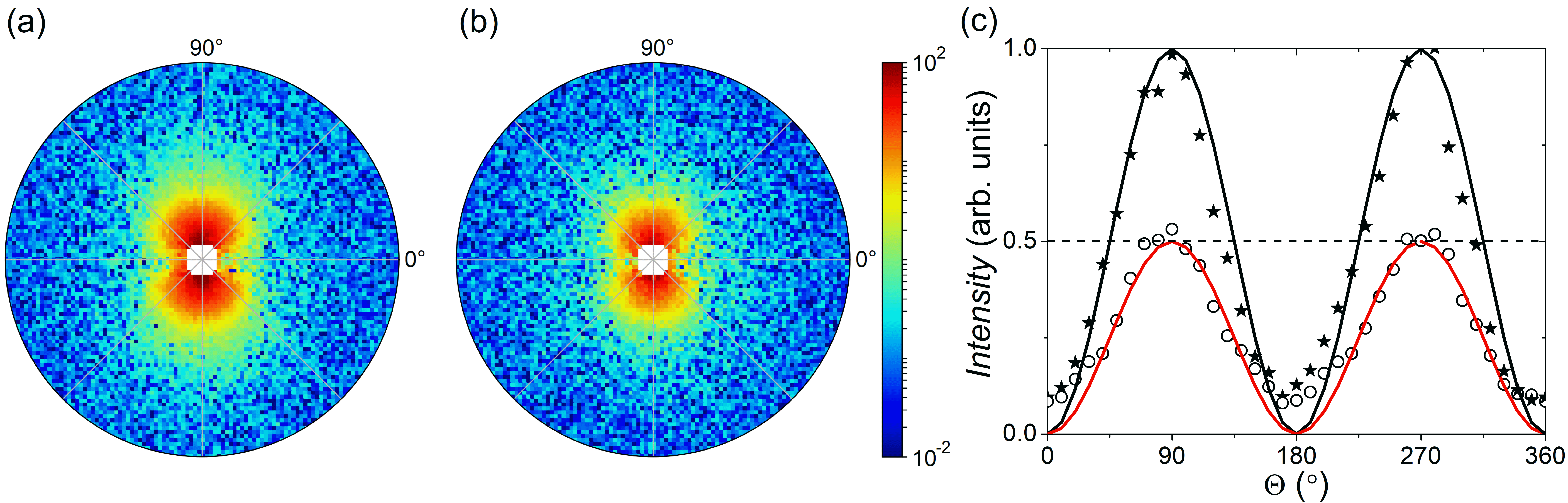}
\caption{\label{Fig4}Residual SANSPOL pattern $I^{-}(\textbf{q})-I^{+}(\textbf{q})=I_\mathrm{cross}(q)\mathrm{sin}^2\Theta$ ($|\textbf{q}|=0.045-0.5\,\mathrm{nm^{-1}}$) in arbitrary units which we detected (a) in the saturated state ($\mu_0H=1\,\mathrm{T}$) and (b) in the remanent state ($\mu_0H=0\,\mathrm{T}$). The homogeneous magnetic field $\textbf{H}$ was applied in horizontal direction along $\Theta=0^\circ$. (c) Intensity integrated over $|\textbf{q}|=0.045-0.5\,\mathrm{nm^{-1}}$ as a function of $\Theta$ of the saturated  (stars) and remanent state (circles) normalized to the maximum of the saturated state. The lines represent $\mathrm{sin}^2\Theta$ (black line) and $1/2\mathrm{sin}^2\Theta$ (red line).} 
\end{figure*}

\begin{figure*}[!h]
\centering  
\includegraphics[width=2\columnwidth]{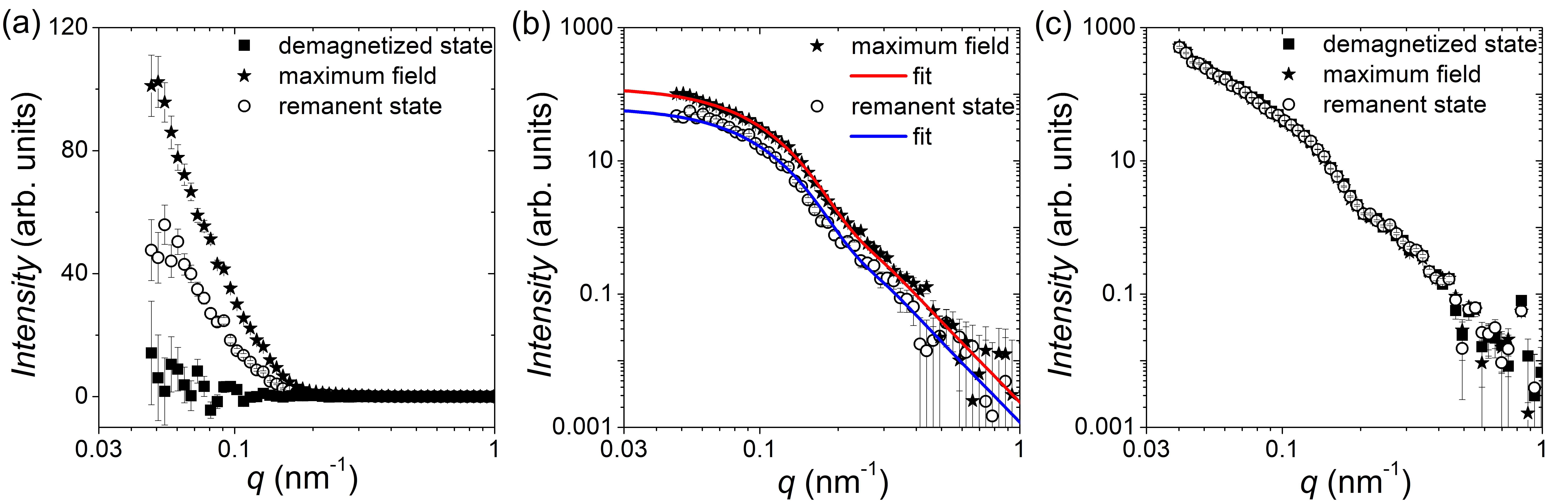}
\caption{\label{Fig5}(a) Nuclear-magnetic cross-term $I_\mathrm{cross}(q)$ detected at maximum field (i.e. magnetic saturation) and at zero field in the demagnetized state and remanent state, respectively (linear $y$-scale). (b) Fit of $I_\mathrm{cross}(q)$ in the saturated and remanent state with the spherical particle form factor (logarithmic $y$-scale). (c) Cross-section of $I^+(\textbf{q})$ parallel to \textbf{H} in the saturated, demagnetized and remanent state. The scattering intensity of the bacteria without magnetosomes is subtracted.} 
\end{figure*}

Fig.\,\ref{Fig2}(a) displays the isothermal magnetization curve of the powder of the freeze-dried bacteria, normalized to the magnetic moment measured in saturation ($m_\mathrm{S}\approx 2\cdot 10^{-5}\,\mathrm{Am^2}$).
In this powder it is safe to assume that the bacteria and hence the 1D chains were randomly oriented (i.e., isotropic ensemble).
The curve is fully saturated at around 0.3\,T and exhibits a hysteresis with a coercive field of $\mu_0H_\mathrm{c}=22\,\mathrm{mT}$ and a normalized remanence of $m_\mathrm{r}/m_\mathrm{S}=0.45$.
This is close to the expected value of 0.5 for an isotropic ensemble of single-domain nanoparticles with uniaxial anisotropy (i.e., Stoner-Wohlfarth particles) \cite{Wohlfarth1958}.
The slightly reduced value 0.45 can be attributed to the small particles at the end of the chains which are superparamagnetic \cite{kumari2014distinguishing}.
The good agreement with the Stoner-Wohlfarth model indicates that during the magnetization process (i) a mechanical particle rotation and (ii) dipolar interactions are negligible, because both would reduce the remanence significantly \cite{bender2013determination,bender2014magnetization,bender2015influence}.

The fact that the magnetosomes do not physically rotate implies a strong mechanical coupling between the magnetosomes and the surrounding bacteria.
At each applied field during the quasi-static magnetization curve the magnetic torque $T_\mathrm{mag}$, and the counteracting mechanical torque $T_\mathrm{mech}$, as well as the torque due to the magnetic anisotropy $T_\mathrm{anis}$, are at equilibrium ($T_\mathrm{mag}$=$T_\mathrm{mech}$=$T_\mathrm{anis}$).
The average magnetic torque $T_\mathrm{mag}=\mu_0HM_\mathrm{S}\left\langle V\right\rangle \mathrm{sin}\Phi$ exerted on the magnetosomes can be directly estimated from the magnetization curve.
Here, $\Phi$ is the average angle between the particle moments and the applied field, which is given by $m/m_\mathrm{S}=\mathrm{cos}\Phi$ \cite{bender2013determination,bender2014magnetization}. 
Using for the saturation magnetization the literature value of magnetite $M_\mathrm{S}=0.48\mathrm{MA/m}$ \cite{coey2010magnetism} and for the mean particle volume $\left\langle V\right\rangle\approx 1/8\pi\left\langle D\right\rangle^{3}$ (with $\left\langle D\right\rangle$ being the mean size according to TEM, i.e., 40\,nm) we get for the maximum of the average torque $T_\mathrm{mag}^\mathrm{max}\approx6.5\cdot 10^{-19}\,\mathrm{Nm}$ (at a field strength of around $\mu_0H=0.1$\,T).
From $T_\mathrm{mag}^\mathrm{max}$ we can calculate the corresponding forces applied on the particle surface to $F_\mathrm{mag}^\mathrm{max}=2\cdot T_\mathrm{mag}^\mathrm{max}/\left\langle D\right\rangle\approx 32.5\,\mathrm{pN}$, which is below the rupture forces (40-80\,pN) between actin filaments and actin-binding proteins reported in literature  \cite{ferrer2008measuring,koernig2014probing}.
Although it has to be considered that we have an isotropic ensemble (i.e., a random orientation distribution) and thus for some magnetosomes, the exerted forces will be higher than 32.5\,pN, this explains why a physical rotation of the magnetosomes is inhibited.

To further evaluate the amount of dipolar interactions we performed modified Henkel and FORC measurements. 
Fig. \ref{Fig2}(b) shows the field-dependence of the normalized isothermal remanence magnetization (IRM) and direct current magnetization (DCD) curves, with $M= m/m_\mathrm{r}$. 
Note that $M_\mathrm{IRM}$ starts in the demagnetized state ($m/m_\mathrm{r}=0$), and saturates at the maximum remanent magnetization with $m/m_\mathrm{r}=1$, while $M_\mathrm{DCD}$ starts at 1 and finishes at -1. 
For the ideal case of non-interacting, uniaxial, single-domain nanoparticles the two remanence curves are related according to the Wohlfarth model ($M_\mathrm{DCD} (H)=1-2M_\mathrm{IRM}(H)$) \cite{Wohlfarth1958}. 
However, if inter-particle interactions appear the relation between $M_\mathrm{IRM}$ and $M_\mathrm{DCD}$ is expected to deviate from the Wohlfarth relation, which can be tested \textit{via} the modified Henkel plot:
\begin{equation}\label{eq:2}
\Delta M= M_\mathrm{DCD} - (1-2M_\mathrm{IRM}).
\end{equation}
The sign of $\Delta M$ provides information about the nature of the interactions, and traditionally negative values ($\Delta M<0$) are interpreted as a sign for the presence of dipolar interactions \cite{fearon1990theoretical}. 
For the magnetosome chains (see Fig.~\ref{Fig2}(b)), we find that $\Delta M$ is slightly negative around $\mu_0H_\mathrm{c}$.
However, the deviation is small, at least in comparison to strongly interacting systems \cite{bender2015influence}, and could also stem from disordered spins \cite{de2017remanence}.

A more sensitive approach to evaluate the magnitude of interactions are FORC measurements \cite{Mayergoyz1986, Pike1999}. 
In these curves, the magnetization $M$, is recorded as a function of the applied field, $\mu_0H$, from the so-called reversal field ($\mu_0H_\mathrm{R}$) up to some positive maximum value (here 100\,mT). 
The complete set of FORCs is obtained by varying the reversal field in discrete steps between the maximum and the minimum field (here -100\,mT). 
Fig.\,\ref{Fig3}(a) shows a group of FORCs (80 curves), measured at room temperature, obtained between 100 and -100\,mT in steps of 2.5\,mT, that fills the interior of the major hysteresis loop. 
These measurements give rise to the FORC distribution presented in Figs.\,\ref{Fig3}(b) and (c) (3D representation and contour plot, respectively), which is calculated from the mixed second order derivative of the magnetization \cite{Roberts2000}
\begin{equation}
\rho(H_\mathrm{R},H)=-\frac{1}{2}\frac{\partial^2M}{\partial H_\mathrm{R}\partial H}
\end{equation}
and plotted as a function of the switching field, $H_\mathrm{s}=(H-H_\mathrm{R})/2$, and the local interactions field, $H_\mathrm{B}=(H+H_\mathrm{R})/2$.
The obtained distribution shows an elongated shape which increases rapidly, similar to a step function and decreases exponentially, as expected for randomly oriented, single-domain particles. 
Moreover, the distribution presents a maximum at around $\mu_0H_\mathrm{S}=20\,\mathrm{mT}$, which is consistent with the coercive field of the complete hysteresis loop (see Fig.\,\ref{Fig2}(a)), and the fact that the distribution is highly peaked on the $H_\mathrm{B} = 0$ axis is also characteristic of a non-interacting system. 

Our magnetometry results are thus in good agreement with previous FORC studies \cite{fischer2008ferromagnetic,egli2010detection,kumari2014distinguishing}, and indicate a dominance of a uniaxial magnetic anisotropy and an absence of dipolar interactions. 
However, this does not mean that there are no interactions between the particles within the chains.
As shown in \citet{charilaou2014magnetic} and \citet{koulialias2016competitive}, due to the collinear arrangement of the anisotropy axes the dipolar stray fields results in an additional uniaxial anisotropy contribution along the chain axis (i.e. shape anisotropy) which enhances the total anisotropy of the 1D chains along the chain axis compared to the individual particles.

Thus, we can conclude from FORC, that the 1D chains of the freeze-dried bacteria do not interact with each other.
Another interesting result we can deduce from magnetometry is that in the freeze-dried bacteria the particles do not mechanically rotate even at high fields (i.e. high magnetic torques), which implies a high mechanical coupling between the magnetosomes and the cytoskeleton.
To further evaluate the stability and also the magnetic properties we performed a polarized SANS experiment on the same sample.

Fig.\,\ref{Fig4} shows the 2D patterns of the residual SANS cross-section $I^{-}(\textbf{q})-I^{+}(\textbf{q})=I_\mathrm{cross}(q)\mathrm{sin}^2\Theta$ at an applied field of 1\,T (Fig.\,\ref{Fig4}(a)) and in the remanent state (Fig.\,\ref{Fig4}(b)).
In Fig.\,\ref{Fig4}(c) we plot additionally both scattering intensities integrated over the $q$-range $0.045-0.5\,\mathrm{nm^{-1}}$ as a function of $\Theta$, which nicely shows the expected $\mathrm{sin}^2$-dependency.
By azimuthally integrating the intensity in the sector perpendicular to $\textbf{H}$ ($\Theta=90^\circ$, $\Delta\Theta=10^\circ$), we extracted the 1D nuclear-magnetic cross-term $I_\mathrm{cross}(q)$.
In Fig.\,\ref{Fig5}(a), we plot $I_\mathrm{cross}(q)$ detected at the maximum field of 1\,T, and detected at zero field in the demagnetized state and in the remanent state, respectively.
As can be seen, for the demagnetized state $I_\mathrm{cross}(q)\approx 0$ over the whole $q$-range, which is expected because the cross-term is directly proportional to the sample magnetization in field direction. 
In the saturated and the remanent state, on the other hand, we detect finite values, whereby the absolute values of the remanent state are by a factor of 2 smaller than in saturation (see also Fig.\,\ref{Fig4}(c)).
This implies a reduced remanence of $m_\mathrm{R}/m_\mathrm{S}=0.5$ (here $m_\mathrm{R}$ is the remanent moment and $m_\mathrm{S}$ the measured moment in saturation), which is the expected value for an isotropic ensemble of single-domain particles with uniaxial anisotropy (i.e. Stoner-Wohlfarth particles) \cite{Wohlfarth1958}, and in good agreement with our magnetization curve (Fig.\,\ref{Fig2}(a)).

We can assume that the nuclear scattering cross-section $\widetilde{N}$ of the bacteria is dominated by the particles.
Thus, in case of a homogeneous magnetization, i.e., single-domain particles with $\widetilde{N}\propto\widetilde{M}_z$, we can write at first approximation for the cross-term $I_\mathrm{cross}(q)\propto F^2(q)$, where $F^2(q)$ is the particle form factor, for which we will use here the spherical one \cite{pedersen1997}.
From $I_\mathrm{cross}(q)$ we can thus additionally estimate the size of the particles by fitting the 1D cross-sections with the spherical form factor (Fig.\,\ref{Fig5}(b)).
We assumed a normal size distribution and determined by a global fit of both data sets (i.e., maximum field and remanent state) for the particle diameter a (number-weighted) mean value of 35(2)\,nm and a standard deviation of 13(1)\,nm.
Comparison of the obtained number-weighted distribution with the size histogram derived by TEM shows a good agreement (Fig.\,\ref{Fig1}(b)).

The similarity between the functional form of the cross-term detected at saturation and remanence (Fig.\,\ref{Fig5}(b)) shows, that qualitatively the magnetic cross-section and magnetic particle morphology is not altered by applying a magnetic field.
This confirms that the particles are in fact (at least at first approximation) single-domain particles and that a rotation of the particle moments into the field direction occurs by an internal, coherent rotation of the atomic magnetic moments inside the individual particles.

Fig.\,\ref{Fig5}(c) shows additionally the 1D cross-sections $I(q)$, which we determined by an azimuthal integration of $I^+(\textbf{q})$ in the sector parallel to $\textbf{H}$ ($\Theta=0^\circ$, $\Delta\Theta=10^\circ$).
The scattering intensity of the bacteria without magnetosomes is subtracted such that we can assume that the SANS signal is dominated by the purely nuclear cross-section of the magnetosomes.
We observe that $I(q)$ does not vary with field leading to a perfect overlap of the scattering curves, which shows that $I(q)$ is dominated by the nuclear scattering, and which in turn verifies that the nuclear cross-section is not changed by applying the magnetic fields.
Hence we can conclude that the particle chains within the freeze-dried bacteria remain intact and do not rotate, because otherwise we would expect a significant variation between the field-dependent cross-sections.
Moreover, the arrangement and separation of the particles composing the magnetosome chains do not vary with field, since no change in the structure factor (i.e., the SANS signal in the low $q$-range) is observed.
Therefore, we can conclude that within the freeze-dried bacteria the internal microstructure is not altered by external fields, which makes them a good system to characterize the magnetic properties of the individual magnetosome chains.
This verifies our own observations \textit{via} TEM, where we could observe no deviations from the chain structure in the probed bacteria, as well as published electron holography studies on similar dried bacteria where also no break-ups were detected \cite{dunin1998magnetic,simpson2005magnetic}.

\section{Conclusions}

To summarize, we investigate here the magnetic response of a powder of freeze-dried and immobilized \textit{M. gryphiswaldense} by magnetometry and polarized SANS.
The magnetosomes are around 40\,nm in size and arranged in linear chains inside the bacteria.
We can confirm that the particles are single-domain particles, and that the 1D magnetosome chains inside the bacteria are not magnetically interacting with each other and behave like randomly oriented magnetic nanoparticles with uniaxial anisotropy along the chain axis.
Furthermore, we can conclude that the particles inside the freeze-dried bacteria do not physically rotate in field direction also in presence of large magnetic fields (i.e. large magnetic torques), which implies a high mechanical coupling between the magnetosomes and the cytoskeletion.
As a result the alignment of the particle moments along the magnetic field-direction occurs exclusively by an internal, coherent rotation. 
Finally, we can show that the particle chains remain intact also after application of large fields of up to 1\,T.
This confirms that the freeze-drying of the bacteria results in mechanically stable configurations of the magnetosome chains, and thus the freeze-dried bacteria can be regarded as model samples to study the magnetization behavior of such 1D nanoparticle chains.

\section*{Acknowledgements}
We thank the Science and Technology Facilities Council (STFC) for granting beamtime at the instrument Larmor. 
P. Bender acknowledges financial support from the National Research Fund of Luxembourg (CORE SANS4NCC grant) and L. Marcano acknowledges the Basque Government for her fellowship (POS{\_}2018{\_}1{\_}0070).
This project has received additional funding from the European Commission Framework Programme 7 under grant agreement no 604448 (NanoMag), and the Spanish Government is acknowledged for funding under project MAT2017-83631-C3-R. 



\bibliography{PBenderBib}

\providecommand*{\mcitethebibliography}{\thebibliography}
\csname @ifundefined\endcsname{endmcitethebibliography}
{\let\endmcitethebibliography\endthebibliography}{}
\begin{mcitethebibliography}{57}
\providecommand*{\natexlab}[1]{#1}
\providecommand*{\mciteSetBstSublistMode}[1]{}
\providecommand*{\mciteSetBstMaxWidthForm}[2]{}
\providecommand*{\mciteBstWouldAddEndPuncttrue}
  {\def\EndOfBibitem{\unskip.}}
\providecommand*{\mciteBstWouldAddEndPunctfalse}
  {\let\EndOfBibitem\relax}
\providecommand*{\mciteSetBstMidEndSepPunct}[3]{}
\providecommand*{\mciteSetBstSublistLabelBeginEnd}[3]{}
\providecommand*{\EndOfBibitem}{}
\mciteSetBstSublistMode{f}
\mciteSetBstMaxWidthForm{subitem}
{(\emph{\alph{mcitesubitemcount}})}
\mciteSetBstSublistLabelBeginEnd{\mcitemaxwidthsubitemform\space}
{\relax}{\relax}

\bibitem[Bazylinski and Frankel(2004)]{Bazylinski2004}
D.~A. Bazylinski and R.~B. Frankel, \emph{Nat. Rev. Microbiol.}, 2004,
  \textbf{2}, 217--230\relax
\mciteBstWouldAddEndPuncttrue
\mciteSetBstMidEndSepPunct{\mcitedefaultmidpunct}
{\mcitedefaultendpunct}{\mcitedefaultseppunct}\relax
\EndOfBibitem
\bibitem[Khalil \emph{et~al.}(2013)Khalil, Pichel, Abelmann, and
  Misra]{khalil2013closed}
I.~S. Khalil, M.~P. Pichel, L.~Abelmann and S.~Misra, \emph{Int. J. Rob. Res.},
  2013, \textbf{32}, 637--649\relax
\mciteBstWouldAddEndPuncttrue
\mciteSetBstMidEndSepPunct{\mcitedefaultmidpunct}
{\mcitedefaultendpunct}{\mcitedefaultseppunct}\relax
\EndOfBibitem
\bibitem[Fdez-Gubieda \emph{et~al.}(2013)Fdez-Gubieda, Muela, Alonso,
  Garcia-Prieto, Olivi, Fernandez-Pacheco, and Barandiaran]{Fdez-Gubieda2013}
M.~L. Fdez-Gubieda, A.~Muela, J.~Alonso, A.~Garcia-Prieto, L.~Olivi,
  R.~Fernandez-Pacheco and J.~M. Barandiaran, \emph{ACS Nano}, 2013,
  \textbf{7}, 3297--3305\relax
\mciteBstWouldAddEndPuncttrue
\mciteSetBstMidEndSepPunct{\mcitedefaultmidpunct}
{\mcitedefaultendpunct}{\mcitedefaultseppunct}\relax
\EndOfBibitem
\bibitem[Komeili \emph{et~al.}(2006)Komeili, Li, Newman, and
  Jensen]{komeili2006magnetosomes}
A.~Komeili, Z.~Li, D.~K. Newman and G.~J. Jensen, \emph{Science}, 2006,
  \textbf{311}, 242--245\relax
\mciteBstWouldAddEndPuncttrue
\mciteSetBstMidEndSepPunct{\mcitedefaultmidpunct}
{\mcitedefaultendpunct}{\mcitedefaultseppunct}\relax
\EndOfBibitem
\bibitem[Toro-Nahuelpan \emph{et~al.}(2019)Toro-Nahuelpan, Giacomelli,
  Raschdorf, Borg, Plitzko, Bramkamp, Sch{\"u}ler, and
  M{\"u}ller]{toro2019mamy}
M.~Toro-Nahuelpan, G.~Giacomelli, O.~Raschdorf, S.~Borg, J.~M. Plitzko,
  M.~Bramkamp, D.~Sch{\"u}ler and F.-D. M{\"u}ller, \emph{Nat. Microbiol.},
  2019, \textbf{4}, 1978--1989\relax
\mciteBstWouldAddEndPuncttrue
\mciteSetBstMidEndSepPunct{\mcitedefaultmidpunct}
{\mcitedefaultendpunct}{\mcitedefaultseppunct}\relax
\EndOfBibitem
\bibitem[Orue \emph{et~al.}(2018)Orue, Marcano, Bender, Garc\'{i}a-Prieto,
  Valencia, Mawass, Gil-Cart\'{o}n, Alba~Venero, Honecker, Garc\'{i}a-Arribas,
  Fern\'{a}ndez~Barqu\'{i}n, Muela, and Fdez-Gubieda]{Orue2018}
I.~Orue, L.~Marcano, P.~Bender, A.~Garc\'{i}a-Prieto, S.~Valencia, M.~A.
  Mawass, D.~Gil-Cart\'{o}n, D.~Alba~Venero, D.~Honecker,
  A.~Garc\'{i}a-Arribas, L.~Fern\'{a}ndez~Barqu\'{i}n, A.~Muela and M.~L.
  Fdez-Gubieda, \emph{Nanoscale}, 2018, \textbf{10}, 7407--7419\relax
\mciteBstWouldAddEndPuncttrue
\mciteSetBstMidEndSepPunct{\mcitedefaultmidpunct}
{\mcitedefaultendpunct}{\mcitedefaultseppunct}\relax
\EndOfBibitem
\bibitem[Lee \emph{et~al.}(2011)Lee, Kim, Choi, Park, Kim, Kim, Choi, Lin, Kim,
  Jung,\emph{et~al.}]{lee2011magnetosome}
N.~Lee, H.~Kim, S.~H. Choi, M.~Park, D.~Kim, H.-C. Kim, Y.~Choi, S.~Lin, B.~H.
  Kim, H.~S. Jung \emph{et~al.}, \emph{PNAS}, 2011, \textbf{108},
  2662--2667\relax
\mciteBstWouldAddEndPuncttrue
\mciteSetBstMidEndSepPunct{\mcitedefaultmidpunct}
{\mcitedefaultendpunct}{\mcitedefaultseppunct}\relax
\EndOfBibitem
\bibitem[Guardia \emph{et~al.}(2012)Guardia, Di~Corato, Lartigue, Wilhelm,
  Espinosa, Garcia-Hernandez, Gazeau, Manna, and Pellegrino]{guardia2012water}
P.~Guardia, R.~Di~Corato, L.~Lartigue, C.~Wilhelm, A.~Espinosa,
  M.~Garcia-Hernandez, F.~Gazeau, L.~Manna and T.~Pellegrino, \emph{ACS Nano},
  2012, \textbf{6}, 3080--3091\relax
\mciteBstWouldAddEndPuncttrue
\mciteSetBstMidEndSepPunct{\mcitedefaultmidpunct}
{\mcitedefaultendpunct}{\mcitedefaultseppunct}\relax
\EndOfBibitem
\bibitem[Lak \emph{et~al.}(2018)Lak, Cassani, Mai, Winckelmans, Cabrera,
  Sadrollahi, Marras, Remmer, Fiorito,
  Cremades-Jimeno,\emph{et~al.}]{lak2018fe2+}
A.~Lak, M.~Cassani, B.~T. Mai, N.~Winckelmans, D.~Cabrera, E.~Sadrollahi,
  S.~Marras, H.~Remmer, S.~Fiorito, L.~Cremades-Jimeno \emph{et~al.},
  \emph{Nano Lett.}, 2018, \textbf{18}, 6856--6866\relax
\mciteBstWouldAddEndPuncttrue
\mciteSetBstMidEndSepPunct{\mcitedefaultmidpunct}
{\mcitedefaultendpunct}{\mcitedefaultseppunct}\relax
\EndOfBibitem
\bibitem[Castellanos-Rubio \emph{et~al.}(2019)Castellanos-Rubio, Rodrigo,
  Munshi, Arriortua, Garitaonandia, Martinez-Amesti, Plazaola, Orue, Pralle,
  and Insausti]{castellanos2019outstanding}
I.~Castellanos-Rubio, I.~Rodrigo, R.~Munshi, O.~Arriortua, J.~S. Garitaonandia,
  A.~Martinez-Amesti, F.~Plazaola, I.~Orue, A.~Pralle and M.~Insausti,
  \emph{Nanoscale}, 2019, \textbf{11}, 16635--16649\relax
\mciteBstWouldAddEndPuncttrue
\mciteSetBstMidEndSepPunct{\mcitedefaultmidpunct}
{\mcitedefaultendpunct}{\mcitedefaultseppunct}\relax
\EndOfBibitem
\bibitem[Martinez-Boubeta \emph{et~al.}(2013)Martinez-Boubeta, Simeonidis,
  Makridis, Angelakeris, Iglesias, Guardia, Cabot, Yedra, Estrad{\'e},
  Peir{\'o},\emph{et~al.}]{martinez2013learning}
C.~Martinez-Boubeta, K.~Simeonidis, A.~Makridis, M.~Angelakeris, O.~Iglesias,
  P.~Guardia, A.~Cabot, L.~Yedra, S.~Estrad{\'e}, F.~Peir{\'o} \emph{et~al.},
  \emph{Sci. Rep.}, 2013, \textbf{3}, 1652\relax
\mciteBstWouldAddEndPuncttrue
\mciteSetBstMidEndSepPunct{\mcitedefaultmidpunct}
{\mcitedefaultendpunct}{\mcitedefaultseppunct}\relax
\EndOfBibitem
\bibitem[Sturm \emph{et~al.}(2019)Sturm, Siglreitmeier, Wolf, Vogel, Gratz,
  Faivre, Lubk, B{\"u}chner, Sturm, and C{\"o}lfen]{sturm2019magnetic}
S.~Sturm, M.~Siglreitmeier, D.~Wolf, K.~Vogel, M.~Gratz, D.~Faivre, A.~Lubk,
  B.~B{\"u}chner, E.~V. Sturm and H.~C{\"o}lfen, \emph{Adv. Funct. Mater.},
  2019, \textbf{29}, 1905996\relax
\mciteBstWouldAddEndPuncttrue
\mciteSetBstMidEndSepPunct{\mcitedefaultmidpunct}
{\mcitedefaultendpunct}{\mcitedefaultseppunct}\relax
\EndOfBibitem
\bibitem[Andreu \emph{et~al.}(2019)Andreu, Urtizberea, and
  Natividad]{andreu2019anisotropic}
I.~Andreu, A.~Urtizberea and E.~Natividad, \emph{Nanoscale}, 2019\relax
\mciteBstWouldAddEndPuncttrue
\mciteSetBstMidEndSepPunct{\mcitedefaultmidpunct}
{\mcitedefaultendpunct}{\mcitedefaultseppunct}\relax
\EndOfBibitem
\bibitem[Yuan \emph{et~al.}(2011)Yuan, Xu, and Mueller]{yuan2011one}
J.~Yuan, Y.~Xu and A.~H. Mueller, \emph{Chem. Soc. Rev.}, 2011, \textbf{40},
  640--655\relax
\mciteBstWouldAddEndPuncttrue
\mciteSetBstMidEndSepPunct{\mcitedefaultmidpunct}
{\mcitedefaultendpunct}{\mcitedefaultseppunct}\relax
\EndOfBibitem
\bibitem[Wang \emph{et~al.}(2011)Wang, Yu, Sun, and Chen]{wang2011magnetic}
H.~Wang, Y.~Yu, Y.~Sun and Q.~Chen, \emph{Nano}, 2011, \textbf{6}, 1--17\relax
\mciteBstWouldAddEndPuncttrue
\mciteSetBstMidEndSepPunct{\mcitedefaultmidpunct}
{\mcitedefaultendpunct}{\mcitedefaultseppunct}\relax
\EndOfBibitem
\bibitem[C{\=e}bers(2005)]{cebers2005flexible}
A.~C{\=e}bers, \emph{Curr. Opin. Colloid Interface Sci.}, 2005, \textbf{10},
  167--175\relax
\mciteBstWouldAddEndPuncttrue
\mciteSetBstMidEndSepPunct{\mcitedefaultmidpunct}
{\mcitedefaultendpunct}{\mcitedefaultseppunct}\relax
\EndOfBibitem
\bibitem[Dreyfus \emph{et~al.}(2005)Dreyfus, Baudry, Roper, Fermigier, Stone,
  and Bibette]{dreyfus2005microscopic}
R.~Dreyfus, J.~Baudry, M.~L. Roper, M.~Fermigier, H.~A. Stone and J.~Bibette,
  \emph{Nature}, 2005, \textbf{437}, 862\relax
\mciteBstWouldAddEndPuncttrue
\mciteSetBstMidEndSepPunct{\mcitedefaultmidpunct}
{\mcitedefaultendpunct}{\mcitedefaultseppunct}\relax
\EndOfBibitem
\bibitem[Kiani \emph{et~al.}(2015)Kiani, Faivre, and Klumpp]{kiani2015elastic}
B.~Kiani, D.~Faivre and S.~Klumpp, \emph{New J. Phys.}, 2015, \textbf{17},
  043007\relax
\mciteBstWouldAddEndPuncttrue
\mciteSetBstMidEndSepPunct{\mcitedefaultmidpunct}
{\mcitedefaultendpunct}{\mcitedefaultseppunct}\relax
\EndOfBibitem
\bibitem[Wolf and Birringer(2005)]{wolf2005pattern}
H.~Wolf and R.~Birringer, \emph{J. Appl. Phys.}, 2005, \textbf{98},
  074303\relax
\mciteBstWouldAddEndPuncttrue
\mciteSetBstMidEndSepPunct{\mcitedefaultmidpunct}
{\mcitedefaultendpunct}{\mcitedefaultseppunct}\relax
\EndOfBibitem
\bibitem[Korth \emph{et~al.}(2006)Korth, Keng, Shim, Bowles, Tang, Kowalewski,
  Nebesny, and Pyun]{korth2006polymer}
B.~D. Korth, P.~Keng, I.~Shim, S.~E. Bowles, C.~Tang, T.~Kowalewski, K.~W.
  Nebesny and J.~Pyun, \emph{J. Am. Chem. Soc.}, 2006, \textbf{128},
  6562--6563\relax
\mciteBstWouldAddEndPuncttrue
\mciteSetBstMidEndSepPunct{\mcitedefaultmidpunct}
{\mcitedefaultendpunct}{\mcitedefaultseppunct}\relax
\EndOfBibitem
\bibitem[Nakata \emph{et~al.}(2008)Nakata, Hu, Uzun, Bakr, and
  Stellacci]{nakata2008chains}
K.~Nakata, Y.~Hu, O.~Uzun, O.~Bakr and F.~Stellacci, \emph{Adv. Mater.}, 2008,
  \textbf{20}, 4294--4299\relax
\mciteBstWouldAddEndPuncttrue
\mciteSetBstMidEndSepPunct{\mcitedefaultmidpunct}
{\mcitedefaultendpunct}{\mcitedefaultseppunct}\relax
\EndOfBibitem
\bibitem[Prozorov \emph{et~al.}(2013)Prozorov, Bazylinski, Mallapragada, and
  Prozorov]{prozorov2013novel}
T.~Prozorov, D.~A. Bazylinski, S.~K. Mallapragada and R.~Prozorov, \emph{Mater.
  Sci. Eng. R Rep.}, 2013, \textbf{74}, 133--172\relax
\mciteBstWouldAddEndPuncttrue
\mciteSetBstMidEndSepPunct{\mcitedefaultmidpunct}
{\mcitedefaultendpunct}{\mcitedefaultseppunct}\relax
\EndOfBibitem
\bibitem[Hoell \emph{et~al.}(2004)Hoell, Wiedenmann, Heyen, and
  Sch{\"u}ler]{hoell2004nanostructure}
A.~Hoell, A.~Wiedenmann, U.~Heyen and D.~Sch{\"u}ler, \emph{Physica B}, 2004,
  \textbf{350}, E309--E313\relax
\mciteBstWouldAddEndPuncttrue
\mciteSetBstMidEndSepPunct{\mcitedefaultmidpunct}
{\mcitedefaultendpunct}{\mcitedefaultseppunct}\relax
\EndOfBibitem
\bibitem[Bender \emph{et~al.}(2019)Bender, Z{\'{a}}kutn{\'{a}}, Disch, Marcano,
  Alba~Venero, and Honecker]{bender2019using}
P.~Bender, D.~Z{\'{a}}kutn{\'{a}}, S.~Disch, L.~Marcano, D.~Alba~Venero and
  D.~Honecker, \emph{Acta Crystallogr., Sect. A: Found. Crystallogr.}, 2019,
  \textbf{75}, 766--771\relax
\mciteBstWouldAddEndPuncttrue
\mciteSetBstMidEndSepPunct{\mcitedefaultmidpunct}
{\mcitedefaultendpunct}{\mcitedefaultseppunct}\relax
\EndOfBibitem
\bibitem[Rosenfeldt \emph{et~al.}(2019)Rosenfeldt, Riese, Mickoleit,
  Sch{\"u}ler, and Schenk]{rosenfeldt2019probing}
S.~Rosenfeldt, C.~N. Riese, F.~Mickoleit, D.~Sch{\"u}ler and A.~S. Schenk,
  \emph{Appl. Environ. Microbiol.}, 2019, \textbf{85}, e01513--19\relax
\mciteBstWouldAddEndPuncttrue
\mciteSetBstMidEndSepPunct{\mcitedefaultmidpunct}
{\mcitedefaultendpunct}{\mcitedefaultseppunct}\relax
\EndOfBibitem
\bibitem[Marcano \emph{et~al.}(2017)Marcano, Garc\'ia-Prieto, Mu\~noz,
  Fern\'andez~Barqu\'in, Orue, Alonso, Muela, and Fdez-Gubieda]{Marcano2017}
L.~Marcano, A.~Garc\'ia-Prieto, D.~Mu\~noz, L.~Fern\'andez~Barqu\'in, I.~Orue,
  J.~Alonso, A.~Muela and M.~L. Fdez-Gubieda, \emph{Biochim. Biophys. Acta,
  Gen. Subj.}, 2017, \textbf{1861}, 1507--1514\relax
\mciteBstWouldAddEndPuncttrue
\mciteSetBstMidEndSepPunct{\mcitedefaultmidpunct}
{\mcitedefaultendpunct}{\mcitedefaultseppunct}\relax
\EndOfBibitem
\bibitem[Marcano \emph{et~al.}(2018)Marcano, Mu\~noz, Mart\'in-Rodr\'iguez,
  Orue, Alonso, Garc\'ia-Prieto, Serrano, Valencia, Abrudan,
  Fern\'andez~Barqu\'in, Garc\'ia-Arribas, Muela, and
  Fdez-Gubieda]{marcano2018magnetic}
L.~Marcano, D.~Mu\~noz, R.~Mart\'in-Rodr\'iguez, I.~Orue, J.~Alonso,
  A.~Garc\'ia-Prieto, A.~Serrano, S.~Valencia, R.~Abrudan,
  L.~Fern\'andez~Barqu\'in, A.~Garc\'ia-Arribas, A.~Muela and M.~L.
  Fdez-Gubieda, \emph{J. Phys. Chem. C}, 2018, \textbf{122}, 7541--7550\relax
\mciteBstWouldAddEndPuncttrue
\mciteSetBstMidEndSepPunct{\mcitedefaultmidpunct}
{\mcitedefaultendpunct}{\mcitedefaultseppunct}\relax
\EndOfBibitem
\bibitem[Gandia \emph{et~al.}(2019)Gandia, Gandarias, Rodrigo,
  Robles-Garc{\'i}a, Das, Garaio, Garc{\'i}a, Phan, Srikanth,
  Orue,\emph{et~al.}]{gandia2019unlocking}
D.~Gandia, L.~Gandarias, I.~Rodrigo, J.~Robles-Garc{\'i}a, R.~Das, E.~Garaio,
  J.~{\'A}. Garc{\'i}a, M.-H. Phan, H.~Srikanth, I.~Orue \emph{et~al.},
  \emph{Small}, 2019, \textbf{15}, 1902626\relax
\mciteBstWouldAddEndPuncttrue
\mciteSetBstMidEndSepPunct{\mcitedefaultmidpunct}
{\mcitedefaultendpunct}{\mcitedefaultseppunct}\relax
\EndOfBibitem
\bibitem[K\"ornig \emph{et~al.}(2014)K\"ornig, Dong, Bennet, Widdrat, Andert,
  M\"uller, Sch\"uler, Klumpp, and Faivre]{koernig2014probing}
A.~K\"ornig, J.~Dong, M.~Bennet, M.~Widdrat, J.~Andert, F.~D. M\"uller,
  D.~Sch\"uler, S.~Klumpp and D.~Faivre, \emph{Nano Lett.}, 2014, \textbf{14},
  4653--4659\relax
\mciteBstWouldAddEndPuncttrue
\mciteSetBstMidEndSepPunct{\mcitedefaultmidpunct}
{\mcitedefaultendpunct}{\mcitedefaultseppunct}\relax
\EndOfBibitem
\bibitem[Blondeau \emph{et~al.}(2018)Blondeau, Guyodo, Guyot, Gatel, Menguy,
  Chebbi, Haye, Durand-Dubief, Alphandery, Brayner, and
  Coradin]{blondeau2018magnetic}
M.~Blondeau, Y.~Guyodo, F.~Guyot, C.~Gatel, N.~Menguy, I.~Chebbi, B.~Haye,
  M.~Durand-Dubief, E.~Alphandery, R.~Brayner and T.~Coradin, \emph{Sci. Rep.},
  2018, \textbf{8}, 7699\relax
\mciteBstWouldAddEndPuncttrue
\mciteSetBstMidEndSepPunct{\mcitedefaultmidpunct}
{\mcitedefaultendpunct}{\mcitedefaultseppunct}\relax
\EndOfBibitem
\bibitem[Heyen and Sch\"uler(2003)]{Heyen2003}
U.~Heyen and D.~Sch\"uler, \emph{Appl. Microbiol. Biotechnol.}, 2003,
  \textbf{61}, 536--544\relax
\mciteBstWouldAddEndPuncttrue
\mciteSetBstMidEndSepPunct{\mcitedefaultmidpunct}
{\mcitedefaultendpunct}{\mcitedefaultseppunct}\relax
\EndOfBibitem
\bibitem[Schneider \emph{et~al.}(2012)Schneider, Rasband, and
  Eliceiri]{Schneider2012}
C.~A. Schneider, W.~S. Rasband and K.~W. Eliceiri, \emph{Nat. Methods}, 2012,
  \textbf{9}, 671\relax
\mciteBstWouldAddEndPuncttrue
\mciteSetBstMidEndSepPunct{\mcitedefaultmidpunct}
{\mcitedefaultendpunct}{\mcitedefaultseppunct}\relax
\EndOfBibitem
\bibitem[Roberts \emph{et~al.}(2000)Roberts, Pike, and Verosub]{Roberts2000}
A.~P. Roberts, C.~R. Pike and Verosub, \emph{J. Geophys. Res.}, 2000,
  \textbf{105}, 28461--28475\relax
\mciteBstWouldAddEndPuncttrue
\mciteSetBstMidEndSepPunct{\mcitedefaultmidpunct}
{\mcitedefaultendpunct}{\mcitedefaultseppunct}\relax
\EndOfBibitem
\bibitem[Honecker \emph{et~al.}(2010)Honecker, Ferdinand, D{\"o}brich,
  Dewhurst, Wiedenmann, G{\'o}mez-Polo, Suzuki, and
  Michels]{honecker2010longitudinal}
D.~Honecker, A.~Ferdinand, F.~D{\"o}brich, C.~D. Dewhurst, A.~Wiedenmann,
  C.~G{\'o}mez-Polo, K.~Suzuki and A.~Michels, \emph{Eur. Phys. J. B}, 2010,
  \textbf{76}, 209--213\relax
\mciteBstWouldAddEndPuncttrue
\mciteSetBstMidEndSepPunct{\mcitedefaultmidpunct}
{\mcitedefaultendpunct}{\mcitedefaultseppunct}\relax
\EndOfBibitem
\bibitem[Michels \emph{et~al.}(2011)Michels, Honecker, D{\"o}brich, Dewhurst,
  Wiedenmann, G{\'o}mez-Polo, and Suzuki]{michels2011small}
A.~Michels, D.~Honecker, F.~D{\"o}brich, C.~Dewhurst, A.~Wiedenmann,
  C.~G{\'o}mez-Polo and K.~Suzuki, \emph{Neutron News}, 2011, \textbf{22},
  15--19\relax
\mciteBstWouldAddEndPuncttrue
\mciteSetBstMidEndSepPunct{\mcitedefaultmidpunct}
{\mcitedefaultendpunct}{\mcitedefaultseppunct}\relax
\EndOfBibitem
\bibitem[Disch \emph{et~al.}(2012)Disch, Wetterskog, Hermann, Wiedenmann,
  Vainio, Salazar-Alvarez, Bergstr{\"o}m, and
  Br{\"u}ckel]{disch2012quantitative}
S.~Disch, E.~Wetterskog, R.~P. Hermann, A.~Wiedenmann, U.~Vainio,
  G.~Salazar-Alvarez, L.~Bergstr{\"o}m and T.~Br{\"u}ckel, \emph{New J. Phys.},
  2012, \textbf{14}, 013025\relax
\mciteBstWouldAddEndPuncttrue
\mciteSetBstMidEndSepPunct{\mcitedefaultmidpunct}
{\mcitedefaultendpunct}{\mcitedefaultseppunct}\relax
\EndOfBibitem
\bibitem[Bender \emph{et~al.}(2018)Bender, Fock, Frandsen, Hansen, Balceris,
  Ludwig, Posth, Wetterskog, Bogart, Southern, Szczerba, Zeng, Witte,
  Gr\"uttner, Westphal, Honecker, Gonz\'alez-Alonso, Fern\'andez~Barqu\'in, and
  Johansson]{bender2018relating}
P.~Bender, J.~Fock, C.~Frandsen, M.~F. Hansen, C.~Balceris, F.~Ludwig,
  O.~Posth, E.~Wetterskog, L.~K. Bogart, P.~Southern, W.~Szczerba, L.~Zeng,
  K.~Witte, C.~Gr\"uttner, F.~Westphal, D.~Honecker, D.~Gonz\'alez-Alonso,
  L.~Fern\'andez~Barqu\'in and C.~Johansson, \emph{J. Phys. Chem. C}, 2018,
  \textbf{122}, 3068--3077\relax
\mciteBstWouldAddEndPuncttrue
\mciteSetBstMidEndSepPunct{\mcitedefaultmidpunct}
{\mcitedefaultendpunct}{\mcitedefaultseppunct}\relax
\EndOfBibitem
\bibitem[M{\"u}hlbauer \emph{et~al.}(2019)M{\"u}hlbauer, Honecker, P{\'e}rigo,
  Bergner, Disch, Heinemann, Erokhin, Berkov, Leighton, Eskildsen, and
  Michels]{muhlbauer2019magnetic}
S.~M{\"u}hlbauer, D.~Honecker, {\'E}.~A. P{\'e}rigo, F.~Bergner, S.~Disch,
  A.~Heinemann, S.~Erokhin, D.~Berkov, C.~Leighton, M.~R. Eskildsen and
  A.~Michels, \emph{Rev. Mod. Phys.}, 2019, \textbf{91}, 015004\relax
\mciteBstWouldAddEndPuncttrue
\mciteSetBstMidEndSepPunct{\mcitedefaultmidpunct}
{\mcitedefaultendpunct}{\mcitedefaultseppunct}\relax
\EndOfBibitem
\bibitem[Coey(2010)]{coey2010magnetism}
J.~M. Coey, \emph{Magnetism and Magnetic Materials}, Cambridge University
  Press, 2010\relax
\mciteBstWouldAddEndPuncttrue
\mciteSetBstMidEndSepPunct{\mcitedefaultmidpunct}
{\mcitedefaultendpunct}{\mcitedefaultseppunct}\relax
\EndOfBibitem
\bibitem[Skomski(2003)]{skomski2003nanomagnetics}
R.~Skomski, \emph{J. Phys.: Condens. Matter}, 2003, \textbf{15}, R841\relax
\mciteBstWouldAddEndPuncttrue
\mciteSetBstMidEndSepPunct{\mcitedefaultmidpunct}
{\mcitedefaultendpunct}{\mcitedefaultseppunct}\relax
\EndOfBibitem
\bibitem[Wohlfarth(1958)]{Wohlfarth1958}
E.~P. Wohlfarth, \emph{J. Appl. Phys.}, 1958, \textbf{29}, 595--596\relax
\mciteBstWouldAddEndPuncttrue
\mciteSetBstMidEndSepPunct{\mcitedefaultmidpunct}
{\mcitedefaultendpunct}{\mcitedefaultseppunct}\relax
\EndOfBibitem
\bibitem[Kumari \emph{et~al.}(2014)Kumari, Widdrat, Tompa, Uebe, Sch{\"u}ler,
  P{\'o}sfai, Faivre, and Hirt]{kumari2014distinguishing}
M.~Kumari, M.~Widdrat, {\'E}.~Tompa, R.~Uebe, D.~Sch{\"u}ler, M.~P{\'o}sfai,
  D.~Faivre and A.~M. Hirt, \emph{J. Appl. Phys.}, 2014, \textbf{116},
  124304\relax
\mciteBstWouldAddEndPuncttrue
\mciteSetBstMidEndSepPunct{\mcitedefaultmidpunct}
{\mcitedefaultendpunct}{\mcitedefaultseppunct}\relax
\EndOfBibitem
\bibitem[Bender \emph{et~al.}(2013)Bender, Tsch{\"o}pe, and
  Birringer]{bender2013determination}
P.~Bender, A.~Tsch{\"o}pe and R.~Birringer, \emph{J. Magn. Magn. Mater.}, 2013,
  \textbf{346}, 152--160\relax
\mciteBstWouldAddEndPuncttrue
\mciteSetBstMidEndSepPunct{\mcitedefaultmidpunct}
{\mcitedefaultendpunct}{\mcitedefaultseppunct}\relax
\EndOfBibitem
\bibitem[Bender \emph{et~al.}(2014)Bender, Tsch{\"o}pe, and
  Birringer]{bender2014magnetization}
P.~Bender, A.~Tsch{\"o}pe and R.~Birringer, \emph{J. Magn. Magn. Mater.}, 2014,
  \textbf{372}, 187--194\relax
\mciteBstWouldAddEndPuncttrue
\mciteSetBstMidEndSepPunct{\mcitedefaultmidpunct}
{\mcitedefaultendpunct}{\mcitedefaultseppunct}\relax
\EndOfBibitem
\bibitem[Bender \emph{et~al.}(2015)Bender, Kr{\"a}mer, Tsch{\"o}pe, and
  Birringer]{bender2015influence}
P.~Bender, F.~Kr{\"a}mer, A.~Tsch{\"o}pe and R.~Birringer, \emph{J. Phys. D:
  Appl. Phys.}, 2015, \textbf{48}, 145003\relax
\mciteBstWouldAddEndPuncttrue
\mciteSetBstMidEndSepPunct{\mcitedefaultmidpunct}
{\mcitedefaultendpunct}{\mcitedefaultseppunct}\relax
\EndOfBibitem
\bibitem[Ferrer \emph{et~al.}(2008)Ferrer, Lee, Chen, Pelz, Nakamura, Kamm, and
  Lang]{ferrer2008measuring}
J.~M. Ferrer, H.~Lee, J.~Chen, B.~Pelz, F.~Nakamura, R.~D. Kamm and M.~J. Lang,
  \emph{PNAS}, 2008, \textbf{105}, 9221--9226\relax
\mciteBstWouldAddEndPuncttrue
\mciteSetBstMidEndSepPunct{\mcitedefaultmidpunct}
{\mcitedefaultendpunct}{\mcitedefaultseppunct}\relax
\EndOfBibitem
\bibitem[Fearon \emph{et~al.}(1990)Fearon, Chantrell, and
  Wohlfarth]{fearon1990theoretical}
M.~Fearon, R.~W. Chantrell and E.~P. Wohlfarth, \emph{J. Magn. Magn. Mater.},
  1990, \textbf{86}, 197--206\relax
\mciteBstWouldAddEndPuncttrue
\mciteSetBstMidEndSepPunct{\mcitedefaultmidpunct}
{\mcitedefaultendpunct}{\mcitedefaultseppunct}\relax
\EndOfBibitem
\bibitem[De~Toro \emph{et~al.}(2017)De~Toro, Vasilakaki, Lee, Andersson,
  Normile, Yaacoub, Murray, S\'anchez, Mun\~niz,
  Peddis,\emph{et~al.}]{de2017remanence}
J.~A. De~Toro, M.~Vasilakaki, S.~S. Lee, M.~S. Andersson, P.~S. Normile,
  N.~Yaacoub, P.~Murray, E.~H. S\'anchez, P.~Mun\~niz, D.~Peddis \emph{et~al.},
  \emph{Chem. Mater.}, 2017, \textbf{29}, 8258--8268\relax
\mciteBstWouldAddEndPuncttrue
\mciteSetBstMidEndSepPunct{\mcitedefaultmidpunct}
{\mcitedefaultendpunct}{\mcitedefaultseppunct}\relax
\EndOfBibitem
\bibitem[Mayergoyz(1986)]{Mayergoyz1986}
I.~Mayergoyz, \emph{IEEE Trans. Magn.}, 1986, \textbf{22}, 603--608\relax
\mciteBstWouldAddEndPuncttrue
\mciteSetBstMidEndSepPunct{\mcitedefaultmidpunct}
{\mcitedefaultendpunct}{\mcitedefaultseppunct}\relax
\EndOfBibitem
\bibitem[Pike \emph{et~al.}(1999)Pike, Roberts, and Verosub]{Pike1999}
C.~R. Pike, A.~P. Roberts and K.~L. Verosub, \emph{J. Appl. Phys.}, 1999,
  \textbf{85}, 6660--6667\relax
\mciteBstWouldAddEndPuncttrue
\mciteSetBstMidEndSepPunct{\mcitedefaultmidpunct}
{\mcitedefaultendpunct}{\mcitedefaultseppunct}\relax
\EndOfBibitem
\bibitem[Fischer \emph{et~al.}(2008)Fischer, Mastrogiacomo, L{\"o}ffler,
  Warthmann, Weidler, and Gehring]{fischer2008ferromagnetic}
H.~Fischer, G.~Mastrogiacomo, J.~F. L{\"o}ffler, R.~J. Warthmann, P.~G. Weidler
  and A.~U. Gehring, \emph{Earth Planet. Sci. Lett.}, 2008, \textbf{270},
  200--208\relax
\mciteBstWouldAddEndPuncttrue
\mciteSetBstMidEndSepPunct{\mcitedefaultmidpunct}
{\mcitedefaultendpunct}{\mcitedefaultseppunct}\relax
\EndOfBibitem
\bibitem[Egli \emph{et~al.}(2010)Egli, Chen, Winklhofer, Kodama, and
  Horng]{egli2010detection}
R.~Egli, A.~P. Chen, M.~Winklhofer, K.~P. Kodama and C.-S. Horng,
  \emph{Geochem. Geophys. Geosyst.}, 2010, \textbf{11}, Q01Z11\relax
\mciteBstWouldAddEndPuncttrue
\mciteSetBstMidEndSepPunct{\mcitedefaultmidpunct}
{\mcitedefaultendpunct}{\mcitedefaultseppunct}\relax
\EndOfBibitem
\bibitem[Charilaou \emph{et~al.}(2014)Charilaou, Kind, Garc{\'\i}a-Rubio,
  Sch{\"u}ler, and Gehring]{charilaou2014magnetic}
M.~Charilaou, J.~Kind, I.~Garc{\'\i}a-Rubio, D.~Sch{\"u}ler and A.~U. Gehring,
  \emph{Appl. Phys. Lett.}, 2014, \textbf{104}, 112406\relax
\mciteBstWouldAddEndPuncttrue
\mciteSetBstMidEndSepPunct{\mcitedefaultmidpunct}
{\mcitedefaultendpunct}{\mcitedefaultseppunct}\relax
\EndOfBibitem
\bibitem[Koulialias \emph{et~al.}(2016)Koulialias, Garc{\'\i}a-Rubio, Rahn-Lee,
  Komeili, L{\"o}ffler, Gehring, and Charilaou]{koulialias2016competitive}
D.~Koulialias, I.~Garc{\'\i}a-Rubio, L.~Rahn-Lee, A.~Komeili, J.~F.
  L{\"o}ffler, A.~U. Gehring and M.~Charilaou, \emph{J. Appl. Phys.}, 2016,
  \textbf{120}, 083901\relax
\mciteBstWouldAddEndPuncttrue
\mciteSetBstMidEndSepPunct{\mcitedefaultmidpunct}
{\mcitedefaultendpunct}{\mcitedefaultseppunct}\relax
\EndOfBibitem
\bibitem[Pedersen(1997)]{pedersen1997}
J.~S. Pedersen, \emph{Adv. Colloid Interface Sci.}, 1997, \textbf{70}, 171 --
  210\relax
\mciteBstWouldAddEndPuncttrue
\mciteSetBstMidEndSepPunct{\mcitedefaultmidpunct}
{\mcitedefaultendpunct}{\mcitedefaultseppunct}\relax
\EndOfBibitem
\bibitem[Dunin-Borkowski \emph{et~al.}(1998)Dunin-Borkowski, McCartney,
  Frankel, Bazylinski, P{\'o}sfai, and Buseck]{dunin1998magnetic}
R.~E. Dunin-Borkowski, M.~R. McCartney, R.~B. Frankel, D.~A. Bazylinski,
  M.~P{\'o}sfai and P.~R. Buseck, \emph{Science}, 1998, \textbf{282},
  1868--1870\relax
\mciteBstWouldAddEndPuncttrue
\mciteSetBstMidEndSepPunct{\mcitedefaultmidpunct}
{\mcitedefaultendpunct}{\mcitedefaultseppunct}\relax
\EndOfBibitem
\bibitem[Simpson \emph{et~al.}(2005)Simpson, Kasama, P{\'o}sfai, Buseck,
  Harrison, and Dunin-Borkowski]{simpson2005magnetic}
E.~Simpson, T.~Kasama, M.~P{\'o}sfai, P.~R. Buseck, R.~Harrison and R.~E.
  Dunin-Borkowski, J. Phys. Conf. Ser., 2005, p. 108\relax
\mciteBstWouldAddEndPuncttrue
\mciteSetBstMidEndSepPunct{\mcitedefaultmidpunct}
{\mcitedefaultendpunct}{\mcitedefaultseppunct}\relax
\EndOfBibitem
\end{mcitethebibliography}
\bibliographystyle{rsc}

\end{document}